\documentclass[prl,superscriptaddress,twocolumn,showpacs,preprintnumbers,%
amsmath,amssymb]{revtex4}

\usepackage{graphicx}
\usepackage{dcolumn}
\usepackage{bm}

\begin{document}
\title{Spin-orbital physics in the optical conductivity of quarter-filled
manganites}
\author{Jan Ba{\l}a}
\affiliation{Max-Planck-Institut f\"ur Festk\"orperforschung,
             Heisenbergstrasse 1, D-70569 Stuttgart, Germany} 
\affiliation{Marian Smoluchowski Institute of Physics, Jagellonian University,
             Reymonta 4, PL-30059 Krak\'ow, Poland}
\author{Peter Horsch}
\affiliation{Max-Planck-Institut f\"ur Festk\"orperforschung,
             Heisenbergstrasse 1, D-70569 Stuttgart, Germany}

\date{\today}

\begin{abstract}
Using finite-temperature diagonalization we investigate
the optical conductivity $\sigma(\omega)$ and spin-orbital dynamics 
in the CE phase
of half-doped manganites. We find $\sigma(\omega)$ characterized 
by a broad spectrum with pronounced optical gap due to 
charge ordering induced by Coulomb and further neighbor Jahn-Teller
interactions. With increasing temperature 
the conductivity shows a significant change over a wide energy range with 
a characteristic shift towards lower frequencies. 
In the low temperature CE phase we observe in-gap absorption due to combined 
orbiton-spin excitations. 
\end{abstract}
\maketitle

\begin{figure}
\includegraphics[width=7cm]{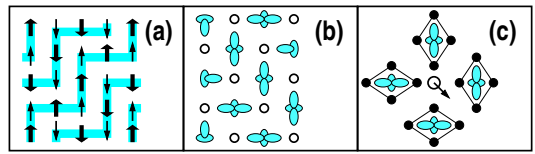}
\caption{\label{fig:cephase}
(Color online)
Sketch of the AF-CE spin, orbital, and lattice order. 
In (a) large (small) spins indicate sites with (without) an $e_g$ electron,
respectively. Shading indicates
direction of DE carrier propagation.
In (b) full symbols represent occupied $e_g$ orbitals
and in (c) full circles indicate the JT-shifted O ions, and the
arrow the displacement of  Mn$^{4+}$.}
\end{figure}

The large variation of physical properties of manganites as function
of doping or temperature originates from a complex interplay of spin,
orbital and charge degrees of freedom, as well as the interaction with
the lattice \cite{Tok00,Tok03}. An important example for the control
of magnetic order due to the orbital degree of freedom is the 
CE phase at quarter filling.
Here ferromagnetic (FM) zig-zag chains are staggered 
antiferromagnetically and the  occupied $e_g$ orbitals are oriented along FM 
bonds in the $(a,b)$ planes \cite{Goo55} [Fig.~\ref{fig:cephase}(a,b)].
Although the structure reflects the cooperative action of 
antiferromagnetic (AF)
superexchange and FM double-exchange (DE), \textit{ these interactions alone}
do not guarantee its stability \cite{Kho01,Bal03}. 
The AF-CE phase was observed in 
cubic (Nd,Pr)$_{1/2}$(Sr,Ca)$_{1/2}$MnO$_3$ \cite{Tok00}
and in layered La$_{1/2}$Sr$_{3/2}$MnO$_4$ \cite{Ste96,Mur98}, 
LaSr$_2$Mn$_2$O$_7$ \cite{Kub99} manganites.
Yet CE-type orbital correlations have also been observed 
in the absence of antiferromagnetism, namely
in the FM metallic phase of Nd$_{1/2}$Sr$_{1/2}$MnO$_3$\cite{Geck02},
and also at smaller
hole doping and are believed to be
connected with the mechanism of colossal magnetoresistance \cite{Mor98}.
An important feature of the CE-phase is a large optical gap  
and a broad absorption
maximum at $1$--$2$ eV shifting towards lower frequencies with increasing 
temperature \cite{Ish99,Jun00,Kim02}. 
Remarkably a 
similar spectral shift is observed in photo-excitation experiments, which
reveal ultrafast response times of such Mott insulating 
structures \cite{Tok03,Oga02}.

Our aim is to investigate
the electron dynamics (optical conductivity, elementary excitations)
in the CE spin-orbital state.
It is evident that a real understanding of this phase is only achieved 
if also the excitation spectra
are consistent with  experiment.
The strong variation of  $\sigma(\omega)$ with temperature $T$
was not addressed in previous theoretical work \cite{Sol01,Cuo02}.
Our study employing finite temperature diagonalization is designed to explore
also quantities, such as the orbital excitation spectra, 
that so far cannot be measured
by experiment and investigate their effect on observables like optical 
conductivity and spin excitations. 
Our results for the optical conductivity reproduce for the first time
the experimental trends on a semi-quantitative level, 
and predict orbital excitations in the charge gap of the  
AF-CE phase.

Orbital degeneracy provides the option for the system to lower its
dimensionality, which is essential for the appearance of the 
CE structure \cite{Yun00}.
Our study explores a novel aspect of orbital physics in manganites, namely 
the role of {\it further neighbor} Jahn-Teller (JT) interactions, 
and the consequences on the dynamics.
We find that 
orbital polarization induced by this interaction is crucial
for the appearance of CE-antiferromagnetism at low T,
and leads to a spin response consistent with AF-CE order.

We consider an extended DE model 
with charge order (CO) triggered by Coulomb repulsion and 
further-neighbor
JT-interaction \cite{Bal03}.
AF order along the $c$ axis and  $e_g$ orbital elongation in 
the $a/b$ direction confines the electron dynamics to $(a,b)$ planes and 
thus we consider a two-dimensional (2D) model.
Assuming infinite on-site Coulomb repulsion between 
$e_g$ electrons on the same site ($U\to \infty$) the model has
the form \cite{Bal03}:
%
\begin{eqnarray}
\label{eq:model}
H=&-&\sum_{\langle{\bf ij}\rangle}
\sum_{\xi\zeta,\sigma}
\left(t^{\xi\zeta}_{\bf ij}\tilde{d}^{\dagger}_{{\bf i}\xi\sigma}
\tilde{d}_{{\bf j}\zeta\sigma}
+ \textrm{H.c.}\right)
-J_H\sum_{\langle{\bf i}\rangle }
{\bf S}_{\bf i}^c\cdot {\bf s_i} \nonumber\\
&+&J_{\textrm AF}\sum_{\langle{\bf ij}\rangle}
{\bf S}_{\bf i}^c\cdot{\bf S}_{\bf j}^c
+V\sum_{\langle{\bf ij}\rangle}n_{\bf i}n_{\bf j}
+H_{\textrm{OO}} .
\end{eqnarray}
The first term describes the motion of  $e_g$ electrons
under the constraint that each site can be occupied at most 
by one $e_g$ electron, 
$\tilde{d}^{\dagger}_{{\bf i}\xi\sigma}=d^{\dagger}_{{\bf i}\xi\sigma}
(1 - n_{{\bf i}\xi\bar{\sigma}})
\prod_{\sigma'}(1-n_{{\bf i}\bar{\xi}\sigma'})$. 
Here the index $\bar{\xi}$ ($\bar{\sigma}$) denotes the $e_g$ orbital 
(spin) orthogonal to $\xi$ ($\sigma$), respectively. 
The hopping matrix elements $t^{\xi\zeta}_{\bf ij}$ in the orbital basis
$\{|x\rangle,|z\rangle\}$ with $|x\rangle\sim x^2-y^2$ 
and $|z\rangle\sim 3z^2-r^2$ have the standard form \cite{Mac99}.
The second 
term in Eq.(\ref{eq:model}) stands for Hund's coupling 
between electron spin ${\bf s_i}=\sum_{\xi\sigma\sigma'}
\tilde{d}^{\dagger}_{{\bf i}\xi\sigma}
\vec{\sigma}_{\sigma\sigma'}  \tilde{d}_{{\bf i}\xi\sigma'} $
and the $t_{2g}$ core spin ${\bf S_i^c}$ ($S^c=3/2$).
The third term represents the  AF superexchange ($J_{\textrm AF}>0$) between 
nearest-neighbor $t_{2g}$ spins, while the fourth term is the inter-site 
Coulomb repulsion which favors checkerboard CO ($V>0$). Finally 
\begin{equation}
H_{\textrm{OO}} = 2\kappa'\sum_{\langle{\bf jij'}
\rangle}(1 - n_{\bf i})T_{\bf jj'} ,
\label{eq:Hoo}
\end{equation}
accounts for the cooperative JT-interaction and
describes the orbital-orbital (OO) interaction between further 
Mn$^{3+}$ neighbors mediated by the Mn$^{4+}$ ion in-between \cite{Bal03}
[Fig.~\ref{fig:cephase}(c)].
Hence, $\langle{\bf jij'}\rangle$ denotes three neighboring 
Mn$^{3+}$--Mn$^{4+}$--Mn$^{3+}$ sites along $a$ or $b$ direction while 
the two-site orbital operator, $T_{\bf jj'}$,
in the $\{|x\rangle,|z\rangle\}$ basis has the form
$T_{\bf jj'} = T_{\bf j}^zT_{\bf j'}^z + 3T_{\bf j}^xT_{\bf j'}^x
\mp \sqrt{3}(T_{\bf j}^xT_{\bf j'}^z + T_{\bf j}^zT_{\bf j'}^x)$ 
expressed in terms of pseudospin operators: 
$T^z_{\bf i}=\frac{1}{2}\sum_{\sigma}(n_{{\bf i}x\sigma}-n_{{\bf i}z\sigma})$ 
and 
$T^{x}_{\bf i}=\frac{1}{2}\sum_{\sigma}
\left(\tilde{d}^{\dagger}_{{\bf i}x\sigma}
\tilde{d}_{{\bf i}z\sigma}+\tilde{d}^{\dagger}_{{\bf i}z\sigma}
\tilde{d}_{{\bf i}x\sigma}\right)$. 
This term not only leads to the CE-type orbital 
order but also induces charge alternation on neighboring Mn sites. 
The  interaction $H_{\textrm{OO}}$ originates from a JT-driven displacement
of the Mn$^{4+}$ octahedra as observed in the CE-phase of 
La$_{1/2}$Ca$_{1/2}$MnO$_3$ \cite{Rad97}. 
The nearest-neighbor OO interaction 
$2\kappa\sum T_{\bf ij}$ \cite{Fei99}
is neglected here for 
simplicity as it does not contribute in the case of strong CO.

First we focus on the optical conductivity
which is determined by the current-current correlation function
\begin{equation}
\sigma(\omega)=\frac{1-e^{-\omega/T}}{N\omega} Re \int_0^{\infty}dt
e^{i\omega t}\langle j_x(t)j_x\rangle,
\label{eq:sigma}
\end{equation}
where the $x$-component of the current operator is       
$j_x=i e \sum_{\langle{\bf ij}\rangle}
\sum_{\xi\zeta,\sigma} t^{\xi\zeta}_{\bf ij} \delta^{\bf ij}_x
\left(\tilde{d}^{\dagger}_{{\bf i}\xi\sigma}
\tilde{d}_{{\bf j}\zeta\sigma}-\textrm{H.c.}\right)$
and $\delta^{\bf ij}_x$
denotes the $x$-component of the vector connecting sites {\bf i} and {\bf j}.
We evaluate this Kubo formula
using the exact diagonalization technique for finite 
temperatures developed by Jakli\v{c} and Prelov\v{s}ek \cite{Jak00} for
cuprates and later used also for manganites \cite{Mac99,Bal03}.
In order to make the problem numerically tractable, we simulated 
the core spins by $S^c=1/2$ spins and scale $J_H$ to preserve the 
splitting between high and low spin states. The data reported here was 
calculated for $J_H=15 t$.
Although we exploited the translational symmetry to save storage 
and computation time,  
calculations were restricted to a small $\sqrt{8}\times\sqrt{8}$
cluster as only 
$S^z_{tot}=\sum_{\bf i} S^z_{\bf i}$ subspaces
(${\bf S_i}={\bf S_i}^c+{\bf s_i}$)
can be treated separately,
while a similar symmetry in the orbital sector is absent.

\begin{figure}
\includegraphics[width=7cm]{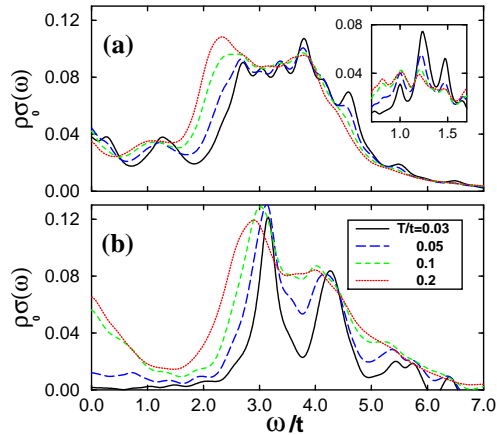}
\caption{\label{fig:cond}
(Color online)
Temperature dependence of the optical conductivity 
$\sigma(\omega)$ obtained in the phase with: (a) AF-CE 
order in the ground state, and (b) a FM-CE  ground state for
$J_{\textrm AF}=0.04t$. Other parameters:
(a) $V=0.2t$, $\kappa'=0.15t$; (b) $V=t$, $\kappa'=0.05t$.
The spectra are generated using a Lorentzian broadening $\Gamma=0.2t$.
Inset: Mid-gap absorption broadened with $\Gamma=0.05t$.}
\end{figure}

The subtle role of $H_{\textrm{OO}}$ is to enhance CE 
type orbital polarization thereby
\textit{supporting the AF decoupling} of FM-chains at low T 
(Fig.~\ref{fig:cephase}a),
while for small $\kappa'$ DE dominates in all directions 
leading to a FM charge and orbital ordered state \cite{Bal03}.
In Fig.~\ref{fig:cond} we show the frequency and temperature dependence
of $\sigma(\omega)$ emerging from ground states with AF-CE
[Fig.~\ref{fig:cond}(a)] and FM-CE [Fig.~\ref{fig:cond}(b)] spin-orbital order,
respectively. A characteristic
feature of $\sigma(\omega)$   is a broad absorption maximum
found for $2t\alt\omega\alt 5t$ which shifts towards lower frequencies by 
$\Delta\omega\simeq 0.5t$ as temperature increases to $T\simeq 0.2t$.
Further increase of temperature leads to the melting of the orbital order 
and the closing of the charge gap ($\omega\alt 2t$) with further 
shift of the high 
frequency broad structure. The shape of $\sigma(\omega)$ and its temperature
evolution is  similar to the experimental data, e.g., for the cubic 
La$_{0.5}$Ca$_{0.5}$MnO$_3$ \cite{Kim02} and layered 
La$_{1/2}$Sr$_{3/2}$MnO$_4$ \cite{Ish99} compound. 
Assuming $t=0.4$ eV \cite{Fei99,Mac99} 
one finds the position of a broad peak structure in
Fig.~\ref{fig:cond} centered  close to 
the experimental value $1.2$ eV measured in  single-layer  
La$_{1/2}$Sr$_{3/2}$MnO$_4$ \cite{Ish99}. 
A subtle test is the absolute scale of  $\sigma(\omega)$
which is controlled by the sum rule 
that links $\sigma(\omega)$
and the kinetic energy of the model \cite{Mac99,Aic02}.
For the maximum of the absorption ($\sigma_{max}\approx 0.1\rho_0^{-1}$)
we find with $\rho_0=\hbar a/e^2$ \cite{Mac99}
and the lattice constant $a\sim 5.5$\AA,   
$\sigma_{max}\approx 500\; (\Omega cm)^{-1}$. Observed values for
$\sigma_{max}$ are $\sim 1000\; (\Omega cm)^{-1}$
in La$_{1/2}$Sr$_{3/2}$MnO$_4$ \cite{Ish99,Jun00} 
and  $\sim 700\; (\Omega cm)^{-1}$
in cubic La$_{0.5}$Ca$_{0.5}$MnO$_3$ \cite{Kim02}.

%
%
The main absorption in Fig.~\ref{fig:cond} describes excitations 
across the charge gap ($\Delta_c\simeq 3(V+2\kappa')$ in case of 
perfect CO) 
which in the orbitally ordered state
involve $e_g$ electron transfer mainly between bridge (ground state) and 
corner (excited state) sites within the same zig-zag chain.
To understand the role of hopping $t$ it is useful
to consider an embedded three-site
model with an $e_g$ electron being either at the bridge ($|B\rangle$)
or at one of the corner ($|C_L\rangle$, $|C_R\rangle$) sites. One finds the
optical transition at $\omega_c\simeq\Delta_c+2t^2/\Delta_c$, between bonding 
$\sim|B\rangle$ and non-bonding $(|C_L\rangle-|C_R\rangle)/\sqrt{2}$ states 
which have different parities. With increasing $T$ magnetic 
excitations are included 
in the thermodynamic average with reduced effective hopping,
whose value is dictated by the DE mechanism for correlated $e_g$ electrons. 
Thus the reduction of the second order term $2t^2/\Delta_c$ for these 
states results in  a spectral 
shift towards lower $\omega_c$. Further reduction of $\omega_c$ 
is caused by melting of orbital and CO leading to gapless orbital 
excitations \cite{Mac99}  and a reduction of $\Delta_c$.

In the FM spin state $\sigma(\omega)$ is characterized by two broad peaks
and a well-developed charge gap [Fig.~\ref{fig:cond} (b)].
Here, spin fluctuations are small and $\sigma(\omega)$ is dominated by 
charge-orbiton excitations, i.e., the pure charge excitation is well separated
from two further peaks corresponding to the additional excitation of 
one and two orbitons. 
On the other hand, in the AF-CE spin 
ordered state spin fluctuations are substantial and inter-site charge transfer 
is accompanied by magnon and orbiton excitations 
leading to a broadened spectrum above 
the charge gap as well as to substantial in-gap absorption. The latter is
completely absent in the FM state at low temperature.

\begin{figure}
\includegraphics[width=7cm]{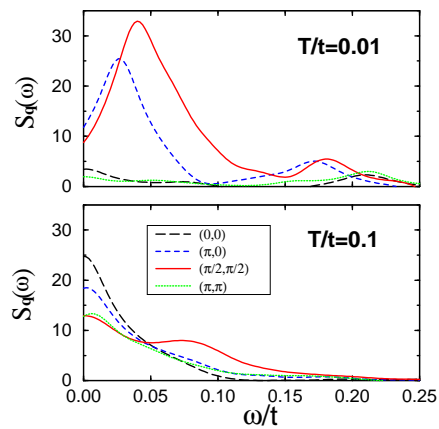}
\caption{\label{fig:SS}
(Color online)
Dynamic spin structure factor $S_{\bf q}(\omega)$ 
in the AF-CE ($T=0.01t$)
and paramagnetic ($T=0.1t$) phase, respectively. Parameters: $V=0.2$,
$\kappa'=0.15t$, $J_{\textrm AF}=0.04t$, $J_H=15t$. Spectra are 
broadened with $\Gamma=0.02t$.}
\end{figure}

To gain a deeper insight into the structure of $\sigma(\omega)$  
in the CE spin-orbital phase, 
we have calculated [using a similar finite-temperature procedure as
for $\sigma(\omega)$] the dynamical spin, orbital, and spin-orbital 
response functions:
\begin{equation}
A_{\bf q}(\omega)=\frac{1}{2\pi}\int_{-\infty}^{\infty}
dt e^{-i\omega t}\langle A^+_{\bf q}A^{-}_{\bf -q}(t)\rangle ,
\end{equation}
where $A_{\bf q}=S_{\bf q}$, $T_{\bf q}$, or $C_{\bf q}$ stand for the spin, 
orbital, or combined spin-orbital operator 
$C_{\bf q}=\sum_{\bf k}S_{\bf k}T_{\bf k+q}$, respectively.
These response functions describe 
the simplest elementary excitations realized in the spin-orbital model.
Strongly broadened collective {\it spin excitations} 
are present at 
low temperatures, but already at $T\simeq 0.1t$ quasi-elastic 
scattering ($\omega\simeq 0$) is found at all momenta (Fig.~\ref{fig:SS}). 
The spin spectrum 
at low temperatures is strongly anisotropic, with pronounced 
precursors of Bragg peaks  
present at momenta $(\pi,0)$ and $(\pi/2,\pi/2)$ at $\omega/t\simeq 0.05$ 
reflecting the spin pattern of the AF-CE phase 
[see Fig.~\ref{fig:cephase}(a)].
Here, spins $S=1$ form two interpenetrating AF superlattices 
(with lattice constant $a=2$) which 
leads to the main maximum in $S_{\bf q}(\omega)$ at ${\bf q}=(\pi/2,\pi/2)$,
while spins $S=1/2$ form one AF lattice in the frame rotated by $45$ 
degrees leading to the smaller maximum in $S_{\bf q}(\omega)$ at 
${\bf q}=(\pi,0)$. Spin waves develop at higher energies $\omega\simeq 0.2t$
due to the competition between AF $t_{2g}$-$t_{2g}$ interactions and DE
mechanism. An accurate assessment of the spin-wave dispersion would, however, 
require the study of larger clusters.
%

\begin{figure}
\includegraphics[width=7cm]{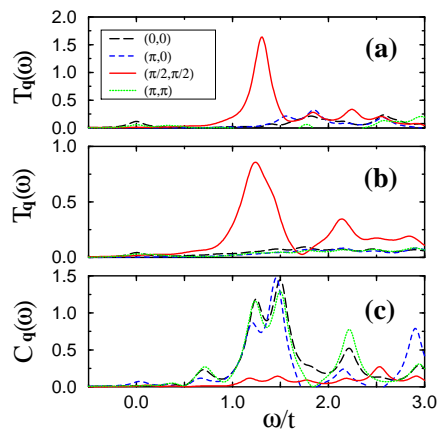}
\caption{\label{fig:TST}
(Color online)
Dynamic structure factor in the 
$\{|x\rangle\pm|z\rangle\}$ basis 
for orbital excitations $T_{\bf q}(\omega)$ 
(a) $T=0.01t$, (b) $T=0.1t$, and (c) combined
spin-orbital excitations $C_{\bf q}(\omega)$ at $T=0.01t$.  
Other parameters as in Fig.~\ref{fig:SS} and $\Gamma=0.1t$.}
\end{figure}

{\it Orbital excitations} are much more robust than the magnons and only
weakly change with increasing temperature from $0.01t$ to $0.1t$.
In the $T_{\bf q}(\omega)$ spectrum [see Fig.~\ref{fig:TST}(a)-(b)]
one finds a pronounced peak
only at momentum ${\bf q}=(\pi/2,\pi/2)$ with $\omega\simeq 1.3t$ at 
low temperatures which broadens with increasing 
temperature and becomes incoherent only for $T\agt t$. 
Less coherent multi-peak 
structure is found at other momenta. The fact that the
orbiton spectrum shows a pronounced peak only at $(\pi/2,\pi/2)$ 
momentum is a consequence of the pseudospin structure of the CE orbital phase 
[see Fig.~\ref{fig:cephase}(b)] which have the same pattern as the $S=1$ 
spin structure [see Fig.~\ref{fig:cephase}(a)].
In the $\{|x\rangle\pm|z\rangle\}$ rotated orbital basis 
the OO term [see Eq.~(\ref{eq:Hoo})] has Ising-like form, 
$T_{\bf ij}\simeq 3T_{\bf i}^zT_{\bf j}^z$, and in the case of perfect 
Mn$^{3+}$--Mn$^{4+}$ CO the energy of orbital excitation is estimated as
$\omega_{orb}\simeq 12\kappa'$. Thus, for $\kappa'=0.15t$ one expects 
$\omega_{orb}\simeq 1.8t$ while in our numerical results
[see Fig.~\ref{fig:TST}(a)] 
the orbiton energy is reduced to  $1.3t$ due to imperfect CO.
%
%
Comparing the spectrum of orbital [Fig.~\ref{fig:TST}(a)] and spin-orbital
[Fig.~\ref{fig:TST}(c)] excitations one finds that collective modes
are formed as well. At energy $\omega\simeq 0.7t$ and different momenta 
one finds a peak well \textit{ below} the orbital mode 
($\omega\simeq 1.3t$) which stands for a bound state \cite{Bri98,Bal01} 
while at $\omega\simeq 1.5t$ a strong maximum is found having a character of
an anti-bound state. This indicates that the orbital excitations, 
although much more robust than the spin ones, play an active role also 
at a lower-energy scale. At higher energies 
$\omega\agt 1.5t$, however,
$C_{\bf q}(\omega)$ and $T_{\bf q}(\omega)$ structure 
factors strongly differ. 

\begin{figure}
\includegraphics[width=8.5cm]{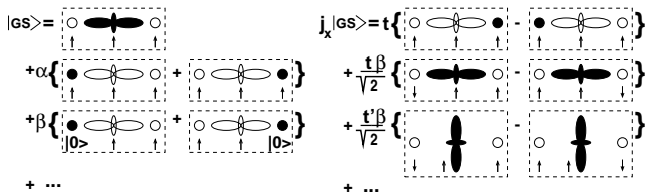}
\caption{\label{fig:proc}
A FM segment showing the dominant terms in the AF-CE ground state
with CO ($\alpha$, $\beta\ll1$) in the limit 
$J_H\to\infty$ (left). When applying
the current operator $j_x$ (right), only the first term would contribute in
the FM case.
The second term $\sim\beta$ in 
$j_x|\textrm{GS}\rangle$ (describing an orbital excitation)
is finite due to  quantum fluctuations in an AF environment.
Here $t$ and $t'$ ($t>t'$) stand for two effective hoppings between corner
and bridge sites, while full (empty) symbols stand for the sites 
occupied (unoccupied) by an $e_g$ electron with total spin (arrows)
$S=1$ ($S=1/2$), and $|0\rangle \equiv |S=1,S^z=0\rangle$.}
\end{figure}

A salient feature of  $\sigma(\omega)$ is the in-gap absorption
($\omega<\Delta_c$) in the AF-CE phase, Fig. 2(a), and the absence of such
absorption for the FM-CE case in Fig. 2(b) at low temperature ($T\alt 0.1t$).  
We argue that the structure at $\omega\simeq t$ is an orbiton-like collective
excitation which becomes infrared-active when  parity is broken by 
AF correlations. This is illustrated schematically 
in Fig.~\ref{fig:proc}. In the CO phase the ground state (GS) is dominated 
by configurations with occupied bridge
sites but hybridized with states reached by inter-site hopping ($\sim t$) or
changed by core spin superexchange ($\sim J_{\rm AF}$). 
As we deal with $d$ orbitals the hopping 
matrix has even parity leading to \textit{ even} 
GS as far as charge distribution 
is concerned. Consequently, the GS changed by 
the current operator is \textit{ odd} with respect to charge and
(neglecting core spins) only processes across the optical gap are allowed
(first line of $j_x|\textrm{GS}\rangle$ in Fig.~\ref{fig:proc}).
In the presence of AF fluctuations 
these selection rules no longer strictly hold, as spin-exchange
leads to an admixture of spin and spin-orbital excitations 
(second and third line of $j_x|\textrm{GS}\rangle$ in Fig.~\ref{fig:proc})
which do not cancel by symmetry.

%
This orbiton-magnon mechanism for the in-gap absorption  
is supported by 
the fact that the intensity of the structure in $\sigma(\omega)$  
at $\omega\simeq t$ 
[see inset in Fig.~\ref{fig:cond}(a)] increases (above the background of other 
excitations) with
decreasing temperature in the spin-ordered case ($T\alt 0.1t$).
The momentum conservation present in optical processes 
can be fulfilled combining 
$T^{\dagger}_{\bf q}$ and $S^z_{\bf -q}$ excitations with
${\bf q}=(\pm\pi/2,\pm\pi/2)$ which are dominant in the respective 
dynamical structure factors [see Fig.~\ref{fig:SS}(a) and 
Fig.~\ref{fig:TST}(a)]. 
The decrease of the static structure factor
$S_{\bf q}=\int_{-\infty}^{\infty}d\omega S_{\bf q}(\omega)$ at 
${\bf q}=(\pm\pi/2,\pm\pi/2)$ with temperature is apparently correlated with 
the suppression of the in-gap absorption in $\sigma(\omega)$.

In summary, optical conductivity and spin-orbital dynamics have been 
investigated in the CE phase of charge-ordered manganites at quarter filling.
Two contributions in our theory determine the charge gap:
(i) elastic forces and (ii) Coulomb interaction. This is important
to explain the recently discovered pico-second switching of optical spectra,
which can be understood by the instantaneous reduction of the Coulomb 
contribution due to the suppression of CO by the ultra-short pulse, 
as well as the fast recovery after the pulse.
We have shown that the cooperative JT coupling
between further Mn neighbors, stabilizing the CE-type correlations 
in our model, can explain
semi-quantitatively the optical data and its temperature dependence in 
half-doped manganites. 
Spin and orbital excitations lead to collective
modes active in  optical experiments, and are responsible for
characteristic differences  between AF- and FM CE phases.

We would like to thank A. M. Ole\'s for fruitful discussions.
J. B. acknowledges the support of the MPI f\"ur Festk\"orperforschung,
Stuttgart and by the Polish State Committee 
of Scientific Research (KBN) of Poland, Project No. 1~P03B~068~26.


\end{document}